# Spin transport in multilayer systems with fully epitaxial NiO thin films

L. Baldrati[1], C. Schneider[1], T. Niizeki[3], R. Ramos[3], J. Cramer[1,2], A. Ross[1,2], E. Saitoh[3,4,5,6], M. Kläui[1,2*]

[1]*Institute of Physics, Johannes Gutenberg-University Mainz, 55128 Mainz, Germany*

[2]*Graduate School of Excellence Materials Science in Mainz, 55128 Mainz, Germany*

[3]*Advanced Institute for Materials Research, Tohoku University, Sendai 980-8577, Japan*

[4] *Institute for Materials Research, Tohoku University, Sendai 980-8577, Japan*

[5]*Advanced Science Research Center, Japan Atomic Energy Agency, Tokai 319-1195, Japan*

[6]*Center for Spintronics Research Network, Tohoku University, Sendai 980-8577, Japan*

*\*Electronic Mail: klaeui@uni-mainz.de*

## ABSTRACT

We report on the generation and transport of thermal spin currents in fully epitaxial $\gamma$-$Fe_2O_3$/NiO(001)/Pt and $Fe_3O_4$/NiO(001)/Pt trilayers. A thermal gradient, perpendicular to the plane of the sample, generates a magnonic spin current in the ferrimagnetic maghemite ($\gamma$-$Fe_2O_3$) and magnetite ($Fe_3O_4$) thin films by means of the spin Seebeck effect. The spin current propagates across the epitaxial, antiferromagnetic insulating NiO layer, before being detected in the Pt layer by the inverse spin Hall effect. The transport of the spin signal is studied as a function of the NiO thickness, temperature and ferrimagnetic material where the spin current is generated. In epitaxial NiO grown on maghemite, the spin Seebeck signal decays exponentially as a function of the NiO thickness, with a spin-diffusion length for thermally-generated magnons of $\lambda_{MSDL} = 1.6 \pm 0.2$ nm, largely independent on temperature. We see no enhancement of the spin current signal as previously reported for certain temperatures and thicknesses of the NiO. In epitaxial NiO grown on magnetite, the temperature-averaged spin diffusion length is $\lambda_{MSDL} = 3.8 \pm 0.3$ nm, and we observe an enhancement of the spin signal when the NiO thickness is 0.8 nm, demonstrating that the growth conditions dramatically affect the spin transport properties of the NiO even for full epitaxial growth. In contrast to theoretical predictions for coherent spin transport, we do not see vastly different spin diffusion lengths between epitaxial and polycrystalline NiO layers.

## 1. Introduction

New types of spintronic devices are envisaged to overcome the limits of current semiconductor-based devices. Magnonic spin currents in insulators do not generate ohmic losses, exhibit a small wavelength (nm), cover a wide frequency range (GHz-THz) and can be used for wave-based computing.[1] In addition, the thermal generation of a magnonic spin current, via the spin Seebeck effect (SSE),[2–4] and its conversion to an electrical voltage may enable small devices and sensors to recover waste heat.[2,5] Yttrium iron garnet (YIG) is the most efficient magnetic insulator for spin transport,[6,7] but iron oxides, in particular magnetite ($Fe_3O_4$)[8,9] and maghemite ($\gamma$-$Fe_2O_3$),[9,10] might be more suitable for the integration with other oxide electronic systems, thanks to the lower cost, the higher abundance of iron with respect to yttrium and the lattice constant matching widely used substrates like MgO(001) and other oxidic compounds. Furthermore, spin currents in antiferromagnetic materials are attracting increased attention due to the absence of stray fields and the typical resonance frequencies in the THz regime.[11,12] Recently, spin transport in insulating antiferromagnets was reported in ferromagnet/antiferromagnet/normal metal (FM/AFM/NM) trilayers, measurable for up to 100 nm of thickness of the inserted AFM,[13–16] and the SSE was observed in pure AFMs driven above the spin-flop transition.[17,18] In the case of FM/AFM/NM trilayers, the spin current generated in the FM propagates through the AFM and is detected in the NM layer by means of the inverse spin Hall effect (ISHE).[19] A number of surprising effects have been found in studies covering spin transmission in polycrystalline AFM layers (especially NiO), such as an enhancement of the spin current upon the introduction of a thin AFM layer. This effect has been reported for some studies on YIG/NiO/Pt systems,[13,14] but was not seen by other groups.[16] Different explanations have been proposed, based on interface effects[20] and coherent magnon transport,[21] and strong dependences on materials quality and crystallinity have been conjectured. However, high quality fully epitaxial stacks have not been probed to date. Spin transport in epitaxial NiO films, having a well-defined structure and a lower density of defects, is thus key to ascertaining the influence of the film quality on the spin transport. Finally, the role of the ferrimagnetic underlayer has not been clarified, since most studies so far have been based on YIG. By varying the ferrimagnetic underlayer, the growth conditions of the multilayer stack change, and it is unclear how this affects the spin transport in the NiO.

In this paper, we report on the thermal generation and transport of magnonic spin currents in epitaxial $\gamma$-$Fe_2O_3$/NiO/Pt trilayers as a function of temperature and thickness of the NiO, to determine the spin transport properties of the antiferromagnet. We compare these results to epitaxial $Fe_3O_4$/NiO/Pt, to demonstrate the importance of the ferrimagnetic underlayer, and thus of the NiO growth, even for fully epitaxial films.

## 2. Experimental

Maghemite ($\gamma$-$Fe_2O_3$) and magnetite ($Fe_3O_4$) are ferrimagnetic insulators with a Curie temperature of 985 K and 860 K, respectively,[9] while nickel oxide (NiO) is an insulating collinear antiferromagnet with a bulk Néel temperature of 523 K,[22] which is reduced for thin films.[23] NiO has a rock salt structure above the Néel temperature (lattice constant 4.176 Å),[24] while maghemite and magnetite comprise spinel structures with lattice constants 8.352 Å[25] and 8.394 Å,[26] respectively. Using MgO(001) substrates, epitaxial growth of layers of these materials can be achieved.[10,27] Epitaxial growth of NiO has been reported on top of magnetite (lattice mismatch 0.5%),[28,29] whilst for maghemite the lattice mismatch is even lower (0.1%). Our samples were grown in a QAM4 sputtering system from ULVAC, with a base pressure of $10^{-5}$ Pa, after pre-annealing the MgO substrates at 800 °C for 2h. The subsequent growth of the $\gamma$-$Fe_2O_3$, $Fe_3O_4$ and NiO films was performed by radio frequency (RF) reactive sputtering of Fe and Ni targets at 430 °C. During sputtering, the same Ar-flow of 15 sccm ($p_{Ar}$ ~ 0.1 Pa) was used for all oxide layers, while the oxygen flow was set at a condition to obtain the best possible epitaxial growth. The $\gamma$-$Fe_2O_3$ and $Fe_3O_4$ layers were grown at an $O_2$-flow of 2.5 sccm and 0.3 sccm, respectively. The $O_2$-flow used for the NiO was either 6.0 sccm or 4.2 sccm, when grown on maghemite and magnetite, respectively. The top Pt layer for detection was deposited in-situ, after cooling the samples down to room temperature in vacuum.

The epitaxial growth of NiO, $Fe_3O_4$ and $\gamma$-$Fe_2O_3$ was checked by x-ray diffraction (XRD) using a Bruker D8 Discover high resolution diffractometer. We here concentrate our analysis first on the $\gamma$-$Fe_2O_3$/NiO system. The $2\theta/\omega$ patterns of thin films grown on MgO(001) substrates, acquired in the symmetric configuration around the reference MgO (002) peak at $2\theta = 42.91°$, are shown in Fig. 1a. The orange curve shows the XRD of a MgO//NiO(50 nm) sample grown at 4.2 sccm of oxygen flow: the NiO(002) peak is situated at $2\theta \sim 43.06°$ ($c_{NiO} \sim 4.20$ Å), implying that the NiO is fully relaxed, while the clear Laue oscillations indicate a high degree of crystallinity along the whole thickness of the film. The relatively high value of the NiO lattice constant may be explained by an excess of oxygen, yielding an increased lattice constant.[30] However, we do not expect the same lattice relaxation to occur for the thinner NiO films (1-10 nm) used for this study. The red curve in Fig. 1a of a MgO//NiO(6 nm)/$\gamma$-$Fe_2O_3$(40 nm) sample shows that the $\gamma$-$Fe_2O_3$ (004) peak is situated at $2\theta \sim 43.73°$ ($c_{\gamma\text{-}Fe2O3} \sim 8.28$ Å), consistent with previous reports.[10] The absence of a $Fe_3O_4$ (004) peak, detected at $2\theta = 43.17°$ in the magnetite samples (see supplementary information), and the lack of a Verwey transition in magnetometry and resistivity measurements (not shown), allows us to exclude the presence of a significant magnetite phase in the samples,[10] while the high saturation magnetization of ~400 kA/m rules out the antiferromagnetic hematite ($\alpha$-$Fe_2O_3$) phase. In order to perform spin Seebeck measurements, we grew MgO//NiO (6 nm)/$\gamma$-$Fe_2O_3$ (40 nm)/NiO (d)/Pt (3.5 nm) stacks, with NiO thicknesses of d = 0, 1, 3, 5, 7, 10 nm, and MgO//NiO (8 nm)/$Fe_3O_4$ (67 nm)/NiO (d)/Pt (3.5 nm) stacks, with NiO thicknesses of d = 0, 0.8, 2.4, 8 nm. The bottom NiO(6, 8 nm) buffer layer is used to avoid Mg diffusion[31] into $\gamma$-$Fe_2O_3$ or $Fe_3O_4$ and compensates the lattice mismatch between the substrate and the iron oxide. This layer does not influence the determination of the spin transport properties of the top NiO, since the thickness of the bottom NiO layer is the same for all the samples of the same set and the thickness of the maghemite (40 nm) and magnetite (67 nm) are higher than their reported spin diffusion lengths.[10,27] The blue curve in Fig. 1a shows the XRD data of MgO//NiO(6 nm)/ γ-Fe$_2$O$_3$(40 nm)/ NiO(50 nm), grown for comparison with the samples used for the measurements based on the spin Seebeck samples. The peaks of γ-Fe$_2$O$_3$ and NiO of the complete stack match the peaks measured in single layer samples, and Laue oscillations for the NiO(002) and the γ-Fe$_2$O$_3$(004) peaks are clearly visible, signaling a high crystalline quality. To obtain a more accurate analysis of the epitaxial growth, we performed symmetric and antisymmetric reciprocal space mapping (RSM) of the same sample, shown in Fig. 1b,c for the reflections in the (002) and ($\bar{1}\bar{1}3$) planes of the MgO substrate, respectively. The alignment of the thin film peaks, along the same h-value as the MgO in the RSM for the ($\bar{1}\bar{1}3$) plane, confirms the expected pseudomorphic growth of NiO(001) and γ-Fe$_2$O$_3$(001) thin films on the MgO(001) substrate, demonstrating the high quality of the deposited epitaxial layers.

## 3. Results and discussion

The MgO//NiO (6 nm)/γ-Fe$_2$O$_3$ (40 nm)/NiO (d)/Pt (3.5 nm) and MgO//NiO (8 nm)/Fe$_3$O$_4$ (67 nm)/NiO (d)/Pt (3.5 nm) samples were cut into 2x10 mm pieces. Two different setups were used for the measurements, one at Mainz university (JGU) was used for maghemite samples, while the setup at Tohoku University (TU) was used for the magnetite ones and both setups yield consistent results, as previously shown.[32] The measurement layout, in the case of the setup at JGU, is depicted in Fig. 2. A constant current, between 10 mA and 20 mA depending on the sample, was applied to the top heater, yielding heating powers of 0.24 W - 0.96 W at room temperature (RT). The resulting perpendicular-to-plane thermal gradient is monitored by the temperature $T_b$ and $T_t$ of the bottom and top of the sample Pt strips, respectively, each measured as a 4-point resistance after the spin Seebeck measurement. Temperature-dependent SSE measurements are performed in a He cryostat equipped with a superconducting magnet and a variable temperature insert. For the detection of the ISHE an in-plane magnetic field between - 1 T and +1 T was aligned along the short edge of the sample, while the transverse voltage was acquired by a Keithley 2182A nanovoltmeter. The detected SSE current $I_{SSE} = V_{ISHE}/R_{Pt}$ was normalized by the temperature difference $\Delta T = T_t - T_b$, cancelling out the variations of the heating power for different samples and ambient temperatures. The setup at Tohoku University (TU), where the $Fe_3O_4$ samples were measured, is based on a Dynacool PPMS cryostat. The sample is stacked between AlN elements with Apiezon grease and the temperature difference is measured by two thermocouples thermally connected to the AlN elements. While the exact temperature gradients at the sample are always difficult to quantitatively ascertain in Spin Seebeck measurements, the temperature and thickness dependences yield reliable information that can be directly compared (see supplementary).[4,33]

In Fig. 3, the estimated spin Seebeck signal $I_{SSE}/\Delta T$, obtained by dividing the spin Seebeck voltage $V_{SSE}$ by the resistance R and the top-bottom temperature difference $\Delta T$ is plotted as a function of temperature for all NiO layers. Dividing the voltage by the resistance allows us to estimate the electrical current induced by the ISHE. Note that the insulating behavior of maghemite allows us to rule out the presence of an Anomalous Nernst Effect (ANE).[10] The detected SSE current exhibits a monotonic increase with temperature, in agreement with previous reports on $\gamma$–$Fe_2O_3$.[10] From the data presented in Fig. 3, we can extract the mean spin diffusion length of thermally generated magnons ($\lambda_{MSDL}$) in NiO by fitting the experimental points at each temperature with a single exponential decay:

$$ln\left(\frac{I_{SSE}}{\Delta T}(d)\right) = ln\left(\frac{I_{SSE}}{\Delta T}(0)\right) - \frac{d}{\lambda_{MSDL}}.$$

The results of the fit are shown in Fig. 4. The increased heating power required to obtain a signal for the thicker (7, 10 nm) NiO samples prevented the acquisition of data points at low temperatures (<100 K). To have a consistent set of data over the whole temperature range, we fitted the thickness dependence using only the samples with NiO layers from 0 nm to 5 nm, obtaining a temperature-averaged spin diffusion length of $\lambda_{MSDL} = 1.6 \pm 0.2$ nm (red curve in Fig. 4), approximately constant across the temperature range investigated. Including the samples of thickness of 7 and 10 nm yields a spin diffusion length that is 10% higher, indicating the existence of long-range spin transport components, possibly associated to multiple magnon modes in the NiO.[34]

To determine the effect of the growth conditions, for the stack including the NiO and the ferrimagnetic material generating the spin current, we next use the iron oxide magnetite ($Fe_3O_4$) as a spin source in MgO//NiO/$Fe_3O_4$/NiO(d)/Pt multilayers, where d = 0.8, 2.4, 8 nm. Note that the ANE is also negligible in $Fe_3O_4$/Pt.[8] The results of these measurements are shown in Fig. 5. The different geometry between the TU and JGU setups yields a different estimate for the $\Delta T$, so that the absolute values of the spin Seebeck coefficient are not directly comparable. However, the thickness and temperature dependences can be compared robustly, as shown in the supplementary information. In the magnetite samples, as was the case for maghemite, the signal monotonically increases as the temperature increases and reaches a plateau at high temperature. The thickness dependence is however very different between maghemite and magnetite, with an enhancement seen at 0.8 nm of NiO thickness in the magnetite series. Excluding the sample without a NiO layer, we obtain in this case $\lambda_{MSDL} = 3.8 \pm 0.3$ nm, which is a factor of two larger than the value obtained for the maghemite-based series. This value is also found to be constant over the probed temperature range T = 100 K – 300 K.

While it was shown that strain can change the properties of NiO, given the polymorphic nature of the multilayer, the in-plane strain experienced by the NiO(001) film is the same for both maghemite and magnetite. This increase of the spin diffusion length is likely related to the different growth conditions of the NiO, where the epitaxial growth of NiO on top of magnetite is obtained at a lower oxygen flow than in maghemite, yielding a longer spin diffusion length. There could also be an effect of the different oxygen coordination between the two ferromagnetic materials, affecting the matching between the magnon modes of the magnetite or maghemite and the NiO, as different modes may have different propagation lengths.[34,35]

Since a detailed study of the spin wave dispersion relation in maghemite is not available, one cannot compare the magnon mode coupling quantitatively at this stage.

The monotonic increase of the spin Seebeck signal as a function of temperature in γ–Fe$_2$O$_3$/NiO/Pt and Fe$_3$O$_4$/NiO/Pt can be attributed to an increase of the magnon population at higher temperatures in the insulating ferro(i)magnets. In this work no decrease in the spin Seebeck signal at higher temperatures is detected in magnetite nor in maghemite, contrary to YIG, where a reduced signal at high temperature stems from a lower magnon spin diffusion length.[4,36] Moreover, we do not see a peak whose temperature depends on the thickness of the NiO, in contrast to reports on polycrystalline NiO and CoO in YIG/NiO/Pt,[14,16] and YIG/CoO/Pt,[15] where an increase in the signal was associated to an increased spin mixing conductance, depending on the transverse spin susceptibility which peaks at the Néel temperature of the AFM layer.[12] The absence of a peak in the probed temperature range is in agreement with the higher Néel temperature expected in epitaxial thin films, with respect to the polycrystalline ones, e.g. Alders *et al.* reported a Néel temperature of 295 K on a thin film of epitaxial NiO ~ 1 nm thick,[23] which is outside of the temperature range studied here. This interpretation also explains why, in contrast to Prakash et al.,[16] we do not observe an increased value of $\lambda_{MSDL}$ for T > 200 K.

Finally, the thickness dependence we observe in maghemite trilayers is characterized by an exponential decay as a function of the NiO thickness, without enhancement of the spin Seebeck signal when a thin epitaxial NiO layer is inserted, consistent with the results obtained for polycrystalline NiO on epitaxial YIG,[16] but in contrast to what was reported for polycrystalline NiO on polycrystalline YIG.[14,16] However, an enhancement of the SSE signal is observed when a NiO layer 0.8 nm thick is included in the stack comprising magnetite and Pt. A possible theoretical explanation put forward relies on coherent spin waves,[21] implying that this enhancement effect should be more pronounced in our single crystalline NiO layers, thanks to the lower defect density expected in epitaxial systems. However, the order of magnitude of the magnon mean spin diffusion length $\lambda_{MSDL}$ in epitaxial NiO on maghemite and magnetite is comparable to the already reported values in polycrystalline NiO on YIG $\lambda_{MSDL} \approx 1 - 10$ nm,[13,14,16] suggesting that it is an intrinsic property of the material and the scattering of the spin waves by grain boundaries is not the dominating factor. Since we do not observe a longer spin diffusion length in our fully epitaxial NiO, and our data in maghemite can be explained by a simple exponential decay law, other non-coherent mechanisms of spin transport must play a role. Moreover, the crystalline orientation of epitaxial NiO(001), allows for spin current transport mainly along the (001) direction and suppresses the spin transmission along possibly more favorable crystallographic directions, available in polycrystalline samples. This might open a path for a future study beyond the scope of the current work, where anisotropic spin diffusion is probed along different crystalline directions. The presence of the signal enhancement when a NiO layer 0.8 nm thick is considered for magnetite samples, not seen in the maghemite-based ones, may be ascribed to the higher spin diffusion length, which shifts the peak of the spin transmission at a measurable NiO thickness. In fact, our observations agree with a model based on incoherent magnon propagation:[20] considering the reduced spin diffusion length of 1.6 nm seen in our NiO films grown in maghemite and using the parameters found in YIG/NiO, the peak in maghemite/NiO is expected at a thickness much lower than the minimum thickness used in this study, which is 1 nm.

## 4. Conclusion

In conclusion, we report a systematic study of the thermal spin transport in high quality, fully epitaxial NiO(001) grown on maghemite and magnetite. The spin Seebeck signal exhibits a monotonic increase as a function of temperature, as compared to the broad peak observed previously for YIG/NiO/Pt. In our single crystalline NiO(001) thin films we find a spin diffusion length that is comparable to the polycrystalline films reported in literature. This implies that the mechanisms that govern the spin diffusion are possibly more complex than what was suggested by previous models based on coherent spin waves.[21] However, we can explain the presence and absence of an enhancement upon the insertion of a NiO layer in magnetite and maghemite, respectively, based on incoherent spin wave models,[20] if the different spin diffusion lengths are taken into account. Our data, in comparison to polycrystalline samples, also suggest that the (001) direction is not necessarily a favorable orientation for long distance spin transport. Finally, we find variations of the spin diffusion length of NiO grown on maghemite and magnetite. This shows that the spin diffusion length in antiferromagnetic insulators is not a universal material constant, but varies with the lattice parameters and the oxygen coordination at the interface. The use of different ferromagnetic underlayers and growth conditions opens up the possibility to tune the spin transport properties of antiferromagnetic materials.

**Acknowledgements**

This work was supported by Deutsche Forschungsgemeinschaft (DFG) SPP 1538 "Spin Caloric Transport," the Graduate School of Excellence Materials Science in Mainz (MAINZ), and the EU project INSPIN (FP7-ICT-2013-X 612759). The authors acknowledge the support of SpinNet (DAAD Spintronics network, project number 56268455), MaHoJeRo (DAAD Spintronics network, project number 57334897), and the DFG (SFB TRR 173 SPIN+X). This work was supported by ERATO "Spin Quantum Rectification Project" (Grant No. JPMJER1402) and Grant-in-Aid for Scientific Research on Innovative Area, "Nano Spin Conversion Science" (Grant No. JP26103005) from JSPS KAKENHI, Japan, and the NEC Corporation. M.K. thanks ICC-IMR at Tohoku University for their hospitality during a visiting researcher stay at the Institute for Materials Research.

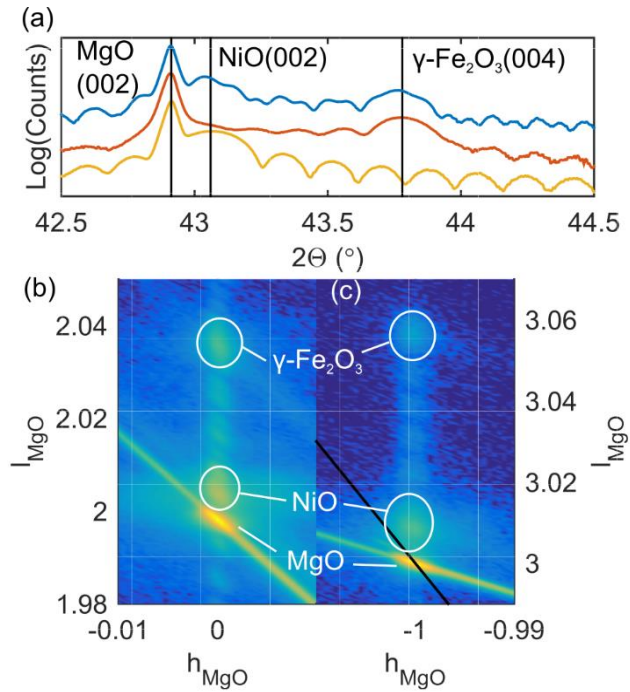

Fig. 1: **(a)** XRD data of NiO(50) (orange line), NiO(6)/γ-Fe$_2$O$_3$(40) (red line) and NiO(6)/γ-Fe$_2$O$_3$(40)/NiO(50) (blue line) stacks grown on MgO substrates (thicknesses in nanometers). The data is vertically shifted for different samples and the vertical black lines are a guide for the eye. **(b)** Symmetric RSM scan of MgO//NiO(6)/γ-Fe$_2$O$_3$(40)/NiO(50) at the (002) MgO peak. The NiO and γ-Fe$_2$O$_3$ peaks are identified by white circles. The vertical, periodic spots are Laue oscillations. **(c)** Antisymmetric RSM scan of MgO//NiO(6)/γ-Fe$_2$O$_3$(40)/NiO(50) at the the ($\bar{1}\bar{1}3$) MgO peak. The black line indicates the expected peak positions from unstrained cubic crystals, while the vertical alignment of the peaks indicates the pseudomorphic growth of the layers.

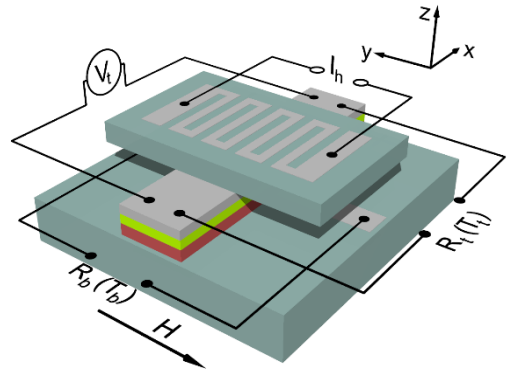

Fig. 2: Schematic of the stack for the spin Seebeck effect used in the JGU setup. The sample is sandwiched between a square sapphire substrate (bottom), with a Pt strip of resistance $R_b$ and temperature $T_b$, and a heater (top) where a heating power $R_h I_h^2$ is injected, generating a temperature gradient. The Pt layer in the sample serves both as a thermometer of resistance $R_t$ and temperature $T_t$, and as a detector of the spin current, using the transverse voltage $V_t$ generated by the ISHE.

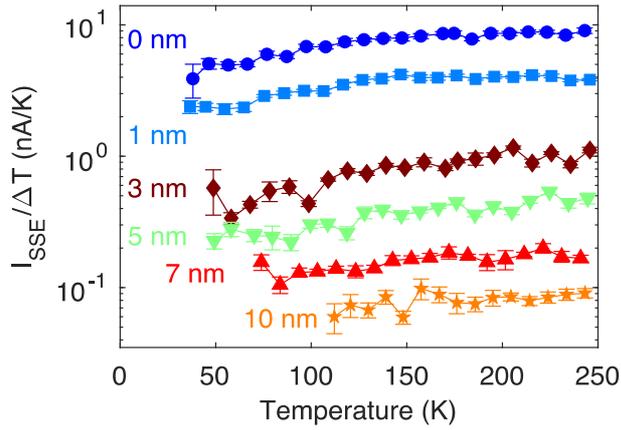

Fig. 3: Temperature dependence of the signal resulting from the thermally generated electrical current ($I_{SSE} = V_{SSE}/R$) in γ-$Fe_2O_3$/NiO/Pt trilayers for different thicknesses (0, 1, 3, 5, 7, 10 nm) of the NiO layer acquired in the setup at JGU. The signal increases up to 150 K, and stays almost constant between 150K and 250K.

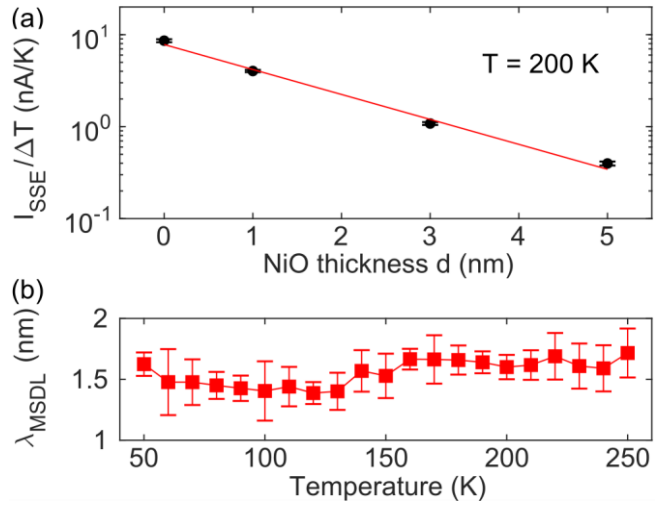

Fig. 4: **(a)** Linear fit of ln($I_{SSE}/\Delta T$) of the maghemite samples at 200K as a function of the thickness of the NiO AFM inserted layer. The fit uses only the four thicknesses (0, 1, 3, 5 nm) for which the full temperature dependence could be obtained, yielding a mean spin diffusion length $\lambda_{MSDL}$ = 1.6 nm. **(b)** Temperature dependence of the mean spin diffusion length. The average value is $\lambda_{MSDL}$ = 1.6 ± 0.2 nm.

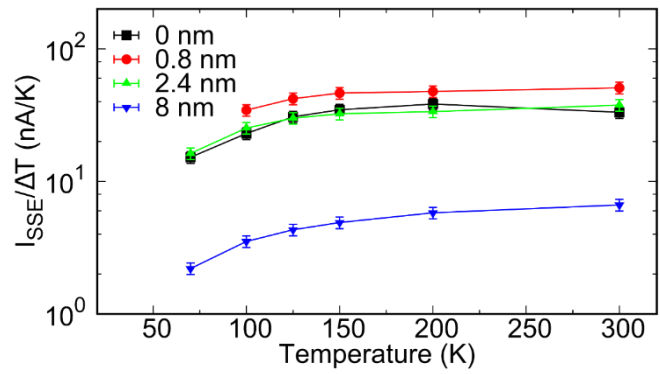

Fig. 5: Temperature dependence of the thermally generated spin current in $Fe_3O_4$/NiO/Pt trilayers for different thicknesses (0, 0.8, 2.4, 8 nm) of the NiO layer acquired in the setup at TU. Note the enhancement of the signal at 0.8 nm thickness of NiO.

# Spin transport in multilayer systems with fully epitaxial NiO thin films – supplementary information

L. Baldrati[1], C. Schneider[1], T. Niizeki[3], R. Ramos[3], J. Cramer[1,2], A. Ross[1,2], E. Saitoh[3,4,5,6], M. Kläui[1,2*]

[1]*Institute of Physics, Johannes Gutenberg-University Mainz, 55128 Mainz, Germany*
[2]*Graduate School of Excellence Materials Science in Mainz, 55128 Mainz, Germany ,*
[3]*Advanced Institute for Materials Research, Tohoku University, Sendai 980-8577, Japan*
[4] *Institute for Materials Research, Tohoku University, Sendai 980-8577, Japan*
[5]*Advanced Science Research Center, Japan Atomic Energy Agency, Tokai 319-1195, Japan*
[6]*Center for Spintronics Research Network, Tohoku University, Sendai 980-8577, Japan*
*\*Electronic Mail: klaeui@uni-mainz.de*

## 1. XRD of magnetite samples

In Fig. S1, the $2\theta/\omega$ x-ray diffraction (XRD) patterns of MgO(001)//NiO(8 nm)/Fe$_3$O$_4$(67 nm)/NiO(d) samples, at various NiO thickness d, are shown. Note the position of the magnetite peak, detected at $2\theta = 43.17°$, compared to the maghemite one, shown in Fig. 1 of the main text, at $2\theta = 43.73°$.[1] When a thin NiO layer is included in the stack on top of the magnetite, the peaks shifts from $2\theta = 43.18°$ to $2\theta = 43.30°$, possibly signaling a strained interface or off-stoichiometry due to the higher oxygen flow used during the growth of NiO.

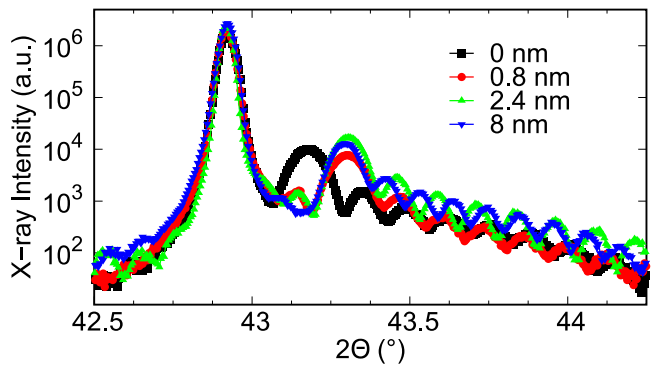

*Figure S1: $2\theta/\omega$ XRD of MgO(001)//NiO(8 nm)/Fe$_3$O$_4$(67 nm)/NiO(d) templates including magnetite, acquired in the symmetric configuration around the reference MgO (002) peak at $2\theta = 42.91°$.*

## 2. Comparison between maghemite and magnetite in different spin Seebeck setups

In order to compare the measurements performed in the setups at Johannes Gutenberg University (JGU) and Tohoku University (TU), we performed spin Seebeck measurements at TU of samples including magnetite (MgO//NiO(8 nm)/Fe$_3$O$_4$(67 nm)/Pt(3.5 nm)) and maghemite (MgO//NiO(6 nm)/γ-Fe$_2$O$_3$(118 nm)/Pt(3.5 nm)), together with measurements at JGU of a MgO//NiO(6 nm)/γ-Fe$_2$O$_3$(40 nm nm)/Pt(3.5 nm) sample. The results are shown in Fig. S2. Note that magnetite (120 nA/K @150 K, TU) is more efficient in terms of generating a spin Seebeck signal than magnetite (35 nA/K @150 K, TU), when they are both measured at the TU setup. The maghemite spin Seebeck signal measured at the TU is 15 times larger than the spin Seebeck signal measured in a maghemite sample 40 nm thick at the JGU (8 nA/K @150 K, JGU), while both signals have the same temperature dependence. The different maghemite thicknesses (40 and 118 nm) cannot explain the increase by a factor of 15 in the spin Seebeck coefficient measured in the two setups, considering that previous work reported few tents of nm as the spin diffusion length in maghemite,[1] and a similar situation is also expected in magnetite. The different results of the measurements arise from the different method used for the determination of the temperature difference. Even if the absolute values of the spin Seebeck coefficient are not directly comparable, the thickness and temperature dependence can be compared robustly even when performed in different setups.

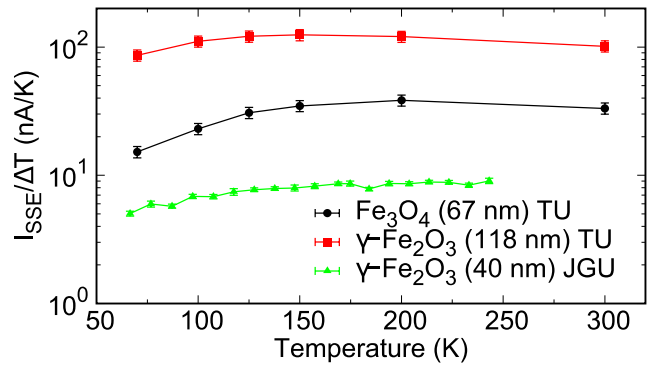

*Figure S2: Comparison of magnetite and maghemite samples, performed with the TU and the JGU setups. Maghemite is more efficient than magnetite in generating a spin transport signal.*

**References**

[1] P. Jiménez-Cavero, I. Lucas, A. Anadón, R. Ramos, T. Niizeki, M.H. Aguirre, P.A. Algarabel, K. Uchida, M.R. Ibarra, E. Saitoh, and L. Morellón, APL Mater. **5**, 026103 (2017).